\documentclass[%
reprint, twocolumn, superscriptaddress,
showpacs,preprintnumbers,
 amsmath,amssymb,
 aps,
pra, nofootinbib
]{revtex4}

\usepackage{graphicx}
\usepackage{dcolumn}
\usepackage{bm}
\usepackage{color} 
\usepackage{CJK}

\def\la{{\langle}}
\def\ra{{\rangle}}

\newcommand{\beq}{\begin{equation}}
\newcommand{\eeq}{\end{equation}}
\newcommand{\beqa}{\begin{eqnarray}}
\newcommand{\eeqa}{\end{eqnarray}}

\begin{document}
\title{Fast transitionless expansions of Gaussian anharmonic traps for cold atoms: bang-singular-bang control}

\author{Xiao-Jing Lu}
\affiliation{Department of Physics, Shanghai University, 200444
Shanghai, People's Republic of China} \affiliation{Departamento de
Qu\'{\i}mica F\'{\i}sica, UPV/EHU, Apdo. 644, 48080 Bilbao, Spain}

\author{Xi Chen}
\email{xchen@shu.edu.cn}
\affiliation{Department of Physics, Shanghai University, 200444
Shanghai, People's Republic of China}

\author{J. Alonso}
\email{alonso@phys.ethz.ch}
\affiliation{Institute for Quantum Electronics, ETH Z\"urich, Otto-Stern-Weg 1,
8093 Z\"urich, Switzerland}

\author{J. G. Muga}
\affiliation{Departamento de Qu\'{\i}mica F\'{\i}sica, UPV/EHU, Apdo.
644, 48080 Bilbao, Spain} \affiliation{Department of Physics,
Shanghai University, 200444 Shanghai, People's Republic of China}

\begin{abstract}
Combining invariant-based inverse engineering, perturbation theory,
and Optimal Control Theory, we design fast, transitionless
expansions of cold neutral atoms or ions in Gaussian anharmonic traps.
Bounding the possible trap
frequencies and using a ``bang-singular-bang'' control we find fast
processes for  a continuum of durations up to a minimum time that
corresponds to a purely bang-bang (stepwise frequency constant)
control.
\end{abstract}
%

\pacs{03.75.-b, 37.10.Gh, 37.10.Ty}

\maketitle

%
%
\section{Introduction}
Cold atoms, neutral or ionized, are manipulated and stored in traps
formed by different electromagnetic field configurations.
Frequently the atom cloud has
to be expanded or compressed, for example to achieve a lower
temperature \cite{temp}, to decrease the velocity spread \cite{Bize}, in cooling cycles
\cite{Salamon}, or simply to adapt the cloud size and facilitate
further operations \cite{duan}. Ideally this processes should be fast, to be able to repeat
them many  times or to avoid decoherence, and should not excite the
system, i.e., they should be ``frictionless'' or
``transitionless''.\footnote{The populations in the
instantaneous bases of the Hamiltonian at initial and final times
should be equal for states that could be connected adiabatically. During
the process, however, transient excitations in the instantaneous
basis are allowed.}
There is currently much interest
in designing fast, transitionless expansions or compressions via
``shortcuts to adiabaticity''
\cite{Masudaprsa,Masudapra,Salamon,prl104,bec,10harmonic-xi,Erik,invariant,Schaff-pra,Schaff-el,Schaff-njp,Onofrio,energy,Adolfo,Bruno,Ronnie2,
resonator,resonatorWu,Andresen,campoel,review}. 
They  speed up adiabatic processes,  reaching the same target states
in shorter times. One of these techniques
is invariant-based inverse engineering
\cite{bec,prl104,energy,invariant}, which has been implemented
experimentally for a cold thermal cloud and for a Bose-Einstein
condensate in a magnetic trap
\cite{Schaff-pra,Schaff-el,Schaff-njp}. In general, it provides
families of solutions to accelerate the dynamics. Optimal
Control Theory (OCT) can be used to select, from  among those families, the ones
that optimize some relevant physical variable, such as the transient
energy \cite{Li,Li-singular,transOCT,Libec}.
%
\begin{figure}[t]
\scalebox{0.5}[0.5]{\includegraphics{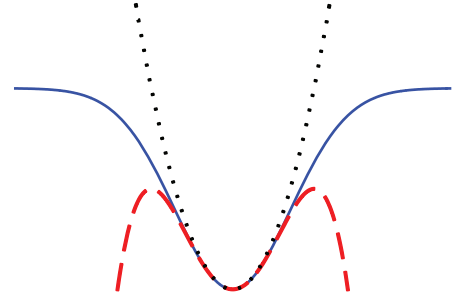}}
\caption{(Color
online)
Potentials for
a Gaussian trap (solid blue), a harmonic trap (dotted black) and an anharmonic trap with a negative quartic term
(dashed red).} \label{fig1}
\end{figure}

Most works devoted to finding shortcuts for trap expansions or compressions considered perfectly
harmonic traps but actual confinements are of course anharmonic. In
this work we shall design, combining invariant-based inverse
engineering, perturbation theory, and OCT, the time dependence of the
trap frequency for 1D confinements with (rather common) Gaussian anharmonicities,
so as to avoid final excitations. Torrontegui \textit{et al.}  investigated the effects of anharmonicity in
three-dimensional cigar-shaped optical traps \cite{Erik}. Here we use a different setting consisting on an effectively 1D
trap with a fixed, tight radial confinement, see for example
\cite{Kino}. We assume on the longitudinal direction $x$ a
superposed
optical potential of Gaussian form \cite{Photo}
\beq\label{gau}
V(x,t)=V_0(t) \left(1-e^{-2x^2/w_0^2}\right),
\eeq
where $V_0(t)$ may be varied by the laser intensity and $w_0$ is the waist.
We also assume a
negligible effect of this potential on the radial confinement and
moderate excitations so that the effective 1D potential takes the
form
\beq
V(x,t)\approx2V_0(t)\!\left(\!\frac{x^2}{w_0^2}-\frac{x^4}{w_0^4}\right)=\frac{m}{2}
\omega^2(t)\!\left(x^2-\frac{x^4}{w_0^2}\!\right)\!,
\eeq
where $\omega^2(t)=4V_0(t)/(mw_0^2)$.
The Gaussian function in Eq. (\ref{gau})
is not unique of optical traps. In particular, the electrostatic potential of an ion trap \cite{ BkMajor} usually resembles a Gaussian function. If we take as an example the trap geometry presented in \cite{Alonso}, the electrode-voltage configurations given therein yield trapping potentials that can be fitted by Gaussian functions with typical coefficients of determination $R^2>0.98$. Linearity of the trapping potential with the voltages applied to the trap electrodes ensures that $w_0$ is independent of $V_0$ for ion traps, too. This makes our results extensible to ions, where trap expansions and compressions
are also of potential interest.
%
%
%
%
%
%
%
\section{Model, invariants and fidelities}
Consider a reduced Hamiltonian of the form
$H(t)=p^{2}/2m+V(x,t)$ for an atom of mass $m$. $H$ can be rewritten
as
\begin{equation}
\label{hami}
H(t)=\underbrace{\frac{p^{2}}{2m}+\frac{1}{2}m\omega^2(t)x^2}_{H_0(t)}
\underbrace{-\frac{1}{2}m\omega^2(t)\frac{x^4}{w^2}}_{V_1(t)}
\end{equation}
in terms of an unperturbed harmonic part and an anharmonic
perturbation $V_{1}$.
This form is not in the family of Lewis-Leach potentials compatible
with quadratic-in-momentum
invariants \cite{LL,review}, so invariant-based
engineering cannot be applied directly, as would be the case for pure
harmonic-oscillator expansions. The alternative route followed here
is to work out first the family of protocols suitable for the purely
harmonic trap; then use perturbation theory to write the fidelity
for the perturbed trap; and, finally, maximize the fidelity (alternatively minimize the
contribution of the anharmonic term to the potential energy).
%
%
%
%
\subsection{Invariant-based inverse engineering for the harmonic oscillator}
A harmonic oscillator with time dependent frequency has the
following invariant \cite{Lewis}:
\begin{equation}
I=\frac{1}{2}\left(\frac{1}{b^{2}}m\omega_{0}^{2}x^{2}+\frac{1}{m}\Pi^{2}\right),
\end{equation}
where $\Pi=bp-m\dot{b}x$ plays the role of a momentum conjugate to
the scaled position $x/b$. The scaling factor $b\equiv b(t)$ satisfies the Ermakov
equation \cite{Lewis}
\begin{equation}
\ddot{b}+\omega^{2}(t)b=\frac{K^{2}}{b^{3}},
\label{ermakov}
\end{equation}
and K is in principle an arbitrary constant, which we fix as
the initial frequency $K=\omega(0)$.
To have $[H_{0}(t),I(t)]=0$ at
$t = 0$ and $t_{f}$, so that these operators  share common eigenfunctions at the boundary times,
we impose the boundary conditions \cite{prl104}
\begin{equation}
\label{boundary}
\begin{split}
b(0)=&1,~~\dot{b}(0)=0,~~\ddot{b}(0)=0,\\
b(t_{f})=&\gamma,~~\dot{b}(t_{f})=0,~~\ddot{b}(t_{f})=0,
\end{split}
\end{equation}
where $\gamma=\left(\omega(0)/\omega(t_f)\right)^{1/2}$. The
dynamical modes are eigenfunctions of $I(t)$ multiplied by
Lewis-Riesenfeld phase factors \cite{Lewis} and have the form
\beqa
&&\!\!\!\!\!\!\la x|\psi_{n}(t)\rangle = \nonumber
\frac{1}{(2^{n}n!b)^{1/2}}\!\left(\!\frac{m\omega(0)}{\pi\hbar}\!\right)^{\!\!1/4}\!\!e^{-i(n+1/2)\omega(0)\!\int_0^{t'}\!\!
\frac{dt''}{b^2(t'')}}
\\ \nonumber
&&\times\exp\!\left[i\frac{m}{2\hbar}\bigg(\frac{\dot{b}}{b}+\frac{i\omega({0})}{b^{2}}\bigg)x^{2}\right]
H_{n}\!\left[\bigg(\frac{m\omega({0})}{\hbar}\bigg)^{\!\!1/2}\!\frac{x}{b}\right].
\\
\eeqa
The solution of the time dependent Schr\"odinger equation with
the Hamiltonian $H_0(t)$ can be expressed as
$|\Psi(t)\rangle=\Sigma_{n}c_{n}|\psi_{n}(t)\rangle$,
where $n=0,1,...$ and the $c_{n}$ are time-independent amplitudes \cite{Lewis}.
\subsection{Perturbation theory}
Let us evaluate the fidelity $F=|\langle\psi_n(t_f)|
\Psi(t_f)\rangle|$, where $\Psi(t)=U(t,0)|\psi_n(0)\ra$ and $U(t,0)$
is the evolution operator for  the Hamiltonian
(\ref{hami}). Using time-dependent perturbation theory,
we approximate
\beqa && \nonumber |\Psi(t_f)\ra=|\psi_n(t_f)\ra -\frac{i}{\hbar}\int_0^{t_f}\!\! dt\,
U_0(t_t,t)V_1(t)|\psi_n(t)\ra
\\\nonumber
&&-\frac{1}{\hbar^2}\!\!\int_0^{t_f}\!\!\!dt\!\!\int_0^t\!\! dt'\,
U_0(t_f,t)V_1(t)U_0(t,t')V_1(t')|\psi_n(t')\ra \!+\!... ,
\\
\eeqa
where $U_0$ is the unperturbed propagator for the invariant-based
trajectory corresponding to the harmonic trap. Using
$U_0(t,t')=\sum_j|\psi_j(t)\ra\la\psi_j(t')|$ we may write, after
cancellation of some terms,
\beq
\label{fidelity}
F=\sqrt{1-\sum_{n\ne n'} |f_{n,n'}^{(1)}|^2},
\eeq
where the first-order transition amplitudes are
\beqa
\label{fnn}
f_{n,n'}^{(1)}&=&\frac{-i}{\hbar}\int_0^{t_f}
dt'\la \psi_n(t')|V_1(t')|\psi_{n'}(t')\ra
\nonumber\\
&=&\frac{i\hbar}{2mw_0^2\omega({0})^{2}}\frac{\alpha_{n,n'}\beta_{n,n'}(t)}{\sqrt{\pi2^{n+n'}n!n'!}},
\label{Fper2}
\\
\alpha_{n,n'}&=&\int_{-\infty}^{\infty} dy\ e^{-y^2}
H_n(y)H_{n'}(y)y^{4},
\eeqa
$H_n$ is the Hermite polynomial, and
\beq
\beta_{n,n'}(t)=\int_0^{t} dt_1\
b^{4}(t_1)\omega^2(t_1)e^{-i(n'-n)\omega({0})\int_{0}^{t_1}
\frac{dt_2}{b^2(t_2)}}. \label{i1}
\eeq
$F$ in Eq. (\ref{fidelity}) is correct to the second order but, since  the
resulting expression is complicated to optimize
we shall use a simpler, approximate approach.
Instead of Eq.
(\ref{fidelity}) we may write the fidelity in terms of diagonal
amplitudes as $|1+f_{n,n}^{(1)}+f_{n,n}^{(2)}+\cdots|$, where
\beq
f_{n,n}^{(2)}=-\frac{1}{2}\sum_n |f_{n,j}^{(1)}|^2.
\eeq
Using the triangular inequality $|x+y|\geq||x|-|y||$,
$F\geq1-|f_{n,n}^{(1)}+f_{n,n}^{(2)}+\cdots|$ and assuming that the
perturbative corrections satisfy $|f_{n,n}^{(1)}|\gg
|f_{n,n}^{(2)}|$, then
\begin{equation}
F\geq 1-|f_{n,n}^{(1)}|.
\end{equation}
In general the right hand side does not provide an accurate approximation for the fidelity but an
approximate  bound
$F_b=1-|f_{n,n}^{(1)}|$,
where, using Eq. (\ref{fnn}) and Ermakov's equation,
\beq
|f_{n,n}^{(1)}|=\frac{3\hbar}{8 m w_0^2}
[(n+1)^2+n^2]\left(t_f-\frac{1}{\omega_{0}^2}
\int_0^{t_f}\ddot{b}b^3 dt\right). \label{modf}
\eeq
The bound may thus be rewritten as
\beqa
F_b
&=& 1- \lambda\int_{0}^{t_{f}}(\omega(0)^2-\ddot{b}b^3) dt,
\label{fb1}
\eeqa
where $\lambda=(3\hbar/4m \omega(0)^2
w_0^2)(n^2+n+1/2).$ Integrating by parts, Eq.
(\ref{fb1}) is finally
\beqa
F_b = 1-\lambda\omega(0)^2t_f-3\lambda\int_{0}^{t_{f}}\dot{b}^2b^2dt.
\label{fb2}
\eeqa
A more rigorous justification for the use of $F_b$ relies on rewriting it as
$F_b=1-\overline{V}_1t_f/\hbar$, where
$$
\overline{V}_1=\frac{1}{t_f}\int_0^{t_f}\langle\psi_n(t)|V_1|\psi_n(t)\rangle dt,
$$
is the time-averaged perturbation energy.
Thus, maximizing $F_b$ amounts to minimize
the time average of the anharmonic perturbation.
Next, we will use OCT to maximize $F_b$
for the ground state $n=0$.
\section{Optimal Control Theory}
If we set $x_{1}=b$, $x_{2}=\dot{b}/\omega_{0}$,
$u(t)=\omega^{2}(t)/\omega_{0}^{2}$ and rescale time according to
$\tau=\omega_{0}t$ (from now on dots are derivatives with respect to
$\tau$),  the Ermakov equation can be replaced by the system
\beqa
\label{x1} \dot{x}_{1}&=&x_{2},
\\
\label{x2} \dot{x}_{2}&=&-ux_{1}+\frac{1}{x^{3}_{1}}.
\eeqa
The optimization goal is to find $u(t)$, constrained by $|u(t)|\leq
\delta$, with $u(0)=1$ and $u(t_f)=1/\gamma^4$, see Eq.
(\ref{boundary}), that minimizes a cost function $J$.
Generally, to minimize the cost function
\begin{equation}
J(u)=\int_0^{\tau_f}g(\textbf{x}(\tau))d\tau,
\end{equation}
the maximum principle states that for the dynamical system
$\dot{\textbf{x}}=\textbf{f}(\textbf{x}(\tau),u)$, the coordinates
of the extremal vector $\textbf{x}(\tau)$ and of the
adjoint state $\textbf{p}(\tau)$ formed by Lagrange multipliers,
fulfill Hamilton's equations for a control Hamiltonian $H_{c}$ \cite{Kirk},
\begin{equation}
\label{xp} \dot{\textbf{x}}=\frac{\partial H_{c}}{\partial
\textbf{p}},~~ \dot{\textbf{p}}=-\frac{\partial H_{c}}{\partial
\textbf{x}},
\end{equation}
where
\begin{equation}
H_c(\textbf{p}(\tau),\textbf{x}(\tau),u)=p_0g(\textbf{x}(\tau))+\textbf{p}^T\cdot
\textbf{f}(\textbf{x}(\tau),u).
\end{equation}
The superscript $T$ here denotes ``transpose''. For $0\leq \tau\leq
\tau_f$, the function $H_c(\textbf{p}(\tau),\textbf{x}(\tau),u)$
attains its maximum at $u=u(\tau)$, and
$H_c(\textbf{p}(\tau),\textbf{x}(\tau),u)=c$, where $c$ is constant.

In the anharmonic trap expansion  we define the cost function
\begin{equation}
J=\int_{0}^{\tau_{f}}x_{1}^{2}x_{2}^{2}d\tau,
\end{equation}
to maximize  $F_b$, see Eq. (\ref{fb2}),
and the control Hamiltonian is
\begin{equation}
 H_{c}=p_{0}x_{1}^{2}x_{2}^{2}+p_{1}x_{2}+p_{2}\left(-ux_{1}+\frac{1}{x^{3}_{1}}\right).
\end{equation}
%
With the control Hamiltonian, Eq. (\ref{xp}) gives the following
costate equations:
\begin{equation}
\label{p}
\begin{split}
\dot{p}_{1}=&-2p_{0}x_{1}x_{2}^{2}+p_{2}u+p_{2}\frac{3}{x_{1}^{4}},\\
\dot{p}_{2}=&-2p_{0}x_{1}^{2}x_{2}-p_{1}.
\end{split}
\end{equation}
\subsection{Bang-bang control}
According to the maximum principle, the control $u(t)$ maximizes the
control Hamiltonian at each time. Note that $H_{c}$ is a linear
function of the control variable $u$. Since $u$ is bounded,
$|u|\leq\delta$, the optimal control that maximizes $H_{c}$ is
determined by the sign of the coefficient of $u$, which is
$-p_{2}x_{1}$. Since $x_{1}>0$, when $p_2\neq 0$, the optimal
control in $(0, \tau_{f})$ is given by
\begin{equation}
u(\tau)=\left\{
         \begin{array}{ll}
           -\delta,~~p_{2}>0 & \hbox{} \\
           \delta,~~p_{2}<0 & \hbox{}
         \end{array}
       \right.,
\end{equation}
so the control sequence is a ``bang-bang'' process \cite{Kirk} with piecewise
constant frequencies,
\begin{equation}
\label{bangbang}
u(\tau)=\left\{
  \begin{array}{ll}
    1, &~~\tau\leq0 \hbox{}, \\
    -\delta, &~~0<\tau<\tau_{1} \hbox{}, \\
    \delta, &~~\tau_1<\tau<\tau_1+\tau_{2} \hbox{}, \\
    1/\gamma^{4},&~~ \tau\geq \tau_{f}\hbox{}.
  \end{array}
  \right.
\end{equation}
At the time boundaries they are given by the boundary conditions, and in
between they saturate the imposed bound in two segments with imaginary and real values.
Since imaginary
frequencies are allowed the bound limits the curvature of the
potential which becomes an anti-trap (repulsive) in the first time segment.
When $u$ is constant and Eqs. (\ref{bangbang}) and (\ref{x2}) are
satisfied, we have
\beq
\label{condition2} x_{2}^{2}+ux_{1}^{2}+\frac{1}{x_{1}^{2}}=c,
\eeq
where $c$ is a constant. Using Eq. (\ref{condition2}) and the
boundary conditions (\ref{boundary}),  $x_1(\tau)$ can
be solved as
\beqa \label{b}
\nonumber
&&x_{1}(\tau)=
\\\nonumber&&\left\{
       \begin{array}{ll}
         \!\!\!\sqrt{\!\frac{\delta-1}{2\delta}+\left(\frac{\delta+1}{2\delta}\right)\cosh2\sqrt{\delta}\tau}, &\tau\in[0,\tau_1] \hbox{}, \\
        \!\!\!\sqrt{\!\frac{\delta\gamma^4+1}{2\delta\gamma^2}\!+\!\left(\frac{\delta\gamma^4-1}{2\delta\gamma^2}\right)\!\cos2\sqrt{\delta}(\tau_f\!-\!\tau)}, &\tau\in[\tau_1,\tau_1\!+\!\tau_2]
        \hbox{},
       \end{array}
     \right.
\\
\eeqa
%
%
%
%
%
%
where $\tau_f = \tau_1+\tau_2$ and
\beqa
\tau_1 &=&
\frac{1}{2{\sqrt{\delta}}}\cosh^{-1}\left[\frac{\delta\gamma^4+1}{\gamma^2(\delta+1)}\right],
\nonumber \\
\tau_2 &=&
\frac{1}{2{\sqrt{\delta}}}\cos^{-1}\left[\frac{\gamma^2(\delta-1)}{\delta\gamma^4-1}\right].
\eeqa
We calculate $\tau_f$ (bang-bang) $= \tau_1+\tau_2$ from the
parameters $\delta$ and $\gamma$,  so $\tau_f$ (bang-bang)
is not arbitrary if they are fixed, as  is usually the case. In
the next section we shall see that arbitrary times larger than
$\tau_f$ (bang-bang) are possible.
\subsection{Bang-singular-bang control}
\subsubsection{Constrained frequency}
When $p_{2}=0$ in some time interval, the maximum principle provides
a priori no information about the optimal (``singular'') control in
this interval \cite{Kirk,Li-singular}. Suppose that $p_{2}=0$ for
$\tau\in[\tau_{1},\tau_1+\tau_{2}]$, then it follows  from Eq.
(\ref{p})
\begin{equation}
\begin{split}
\dot{p}_{1}=&-2p_{0}x_{1}x_{2}^{2},\\
p_{1}=&-2p_{0}x_{1}^{2}x_{2}.
\end{split}
\end{equation}
Therefore
\begin{equation}
\dot{x}_{2}x_{1}+x_{2}^{2}=0.
\end{equation}
Integrating the above equation, we have
\begin{equation}
\label{condition1} x_{2}=\frac{c_1}{x_{1}}.
\end{equation}
Then $x_1(\tau)$ takes the form
\begin{equation}
x_1(\tau)=\sqrt{c_1 \tau+c_{2}},~~\tau\in[\tau_{1},\tau_1+\tau_{2}].
\end{equation}
Using  Eq. (\ref{x2}), the control on the singular point $p_{2}=0$
is given by
\begin{equation}
u_{s}=\frac{1+x_{1}^2 x_2^2}{x_1^4}.
\end{equation}
The  ``bang-singular-bang" control sequence \cite{Kirk,Li-singular} with two intermediate
switchings at  $\tau=\tau_{1}$ and $\tau=\tau_1+\tau_{2}$ is
\begin{equation}
\label{b-s-b} u(t)=\left\{
  \begin{array}{ll}
    1, &~~\tau\leq0 \hbox{}, \\
    -\delta, &~~0<\tau<\tau_{1} \hbox{}, \\
    u_{s}, &~~\tau_{1}<\tau<\tau_1+\tau_{2} \hbox{}, \\
    \delta, &~~\tau_1+\tau_{2}<\tau<\tau_{f} \hbox{}, \\
    1/\gamma^{4},&~~ \tau\geq \tau_{f}\hbox{}.
  \end{array}
\right.
\end{equation}
where $\tau_f=\tau_1+\tau_{2}+\tau_3$.

Using the boundary conditions (\ref{boundary}) and (\ref{b-s-b}),
the function $x_{1}(\tau)$ can be solved as
\beqa \label{b}
\nonumber&&x_{1}(\tau)=
\\
\nonumber&&\!\!\left\{
       \!\begin{array}{ll}
         \!\!\sqrt{\frac{\delta-1}{2\delta}+\left(\frac{\delta+1}{2\delta}\right)\cosh2\sqrt{\delta}\tau}, & \tau\in[0,\tau_1]\hbox{}, \\
         \sqrt{2 c_1 \tau+c_{2}}, &\tau\in[\tau_1,\tau_{1}\!+\!\tau_2] \hbox{}, \\
        \!\sqrt{\frac{\delta\gamma^4+1}{2\delta\gamma^2}+\left(\frac{\delta\gamma^4-1}{2\delta\gamma^2}\right)\cos2\sqrt{\delta}(\tau_f\!-\!\tau)},&t\in[\tau_1\!+\!\tau_2,\tau_f]
        \hbox{}.
       \end{array}
     \right.
\\
\eeqa
%
%
%
%
%
%
%
%
%
From the boundary conditions, Eqs. (\ref{boundary}) and
(\ref{condition2}), we find the trajectory
\begin{equation}
\label{condition2-1} x_{2}^2-\delta x_1^2+\frac{1}{x_1^2}=1-\delta,
\end{equation}
for $0 \leq \tau \leq \tau_1$ and
\begin{equation}
\label{condition2-2}
x_{2}^2+\delta x_1^2+\frac{1}{x_1^2}=\delta
\gamma^2+\frac{1}{\gamma^2},
\end{equation}
for $\tau_1+\tau_2\leq \tau \leq \tau_f$.
Then we solve the first junction point, using Eqs.
(\ref{condition1}) and (\ref{condition2-1}), as
\begin{equation}
\label{x1t1}
x_1^2(\tau_1)=\frac{\delta-1+\sqrt{\delta^2+(4c_1^2+2)\delta+1}}{2\delta},
\end{equation}
and at the second junction we get from Eqs. (\ref{condition1}) and
(\ref{condition2-2})
\begin{equation}
\label{x1t1t2}
x_1^2(\tau_1+\tau_2)=\frac{\delta\gamma^4+1+\sqrt{\delta^2\gamma^8-(4c_1^2+2)\delta\gamma^4+1}}{2\delta\gamma^2}.
\end{equation}

Because of the continuity of the function $x_1(\tau)$, the interval
times $\tau_{1,2,3}$ can be found from Eq. (\ref{b}),
%
\beqa \label{tau}
\tau_1(c_1)&=&\frac{1}{2\sqrt{\delta}}\cosh^{-1}\left[\frac{2\delta
x_1^2(\tau_1)-\delta-1}{\delta+1}\right],
\\\nonumber
\tau_2(c_1)&=&\frac{1}{2c_1}\left[x_1^2(\tau_1+\tau_2)-x_1^2(\tau_1)\right],
\\\nonumber
%
\tau_3(c_1)&=&\frac{1}{2\sqrt{\delta}}\cos^{-1}\left[\frac{2\delta\gamma^2
x_1^2(\tau_1+\tau_2)-\delta\gamma^4-1}{\delta\gamma^4-1}\right].
\eeqa
The final time $\tau_f$ is
\begin{equation}
\tau_f=\tau_1(c_1)+\tau_2(c_1)+\tau_3(c_1),
\end{equation}
which determines the constant $c_1$, and thus $c_2=x_1^2(\tau_1)-2c_1\tau_1$.

\begin{figure}[h]
\scalebox{0.5}[0.5]{\includegraphics{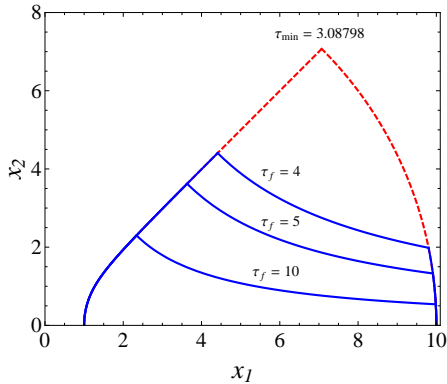}} \caption{(Color
online) Trajectories corresponding to different final times, $\tau_f
= 4, 5, 10$ and $\tau_{min} = 3.08798$. Parameters: $\gamma=10$ and
$\delta=1$. } \label{fig2}
\end{figure}

\begin{figure}[h]
\scalebox{0.5}[0.5]{\includegraphics{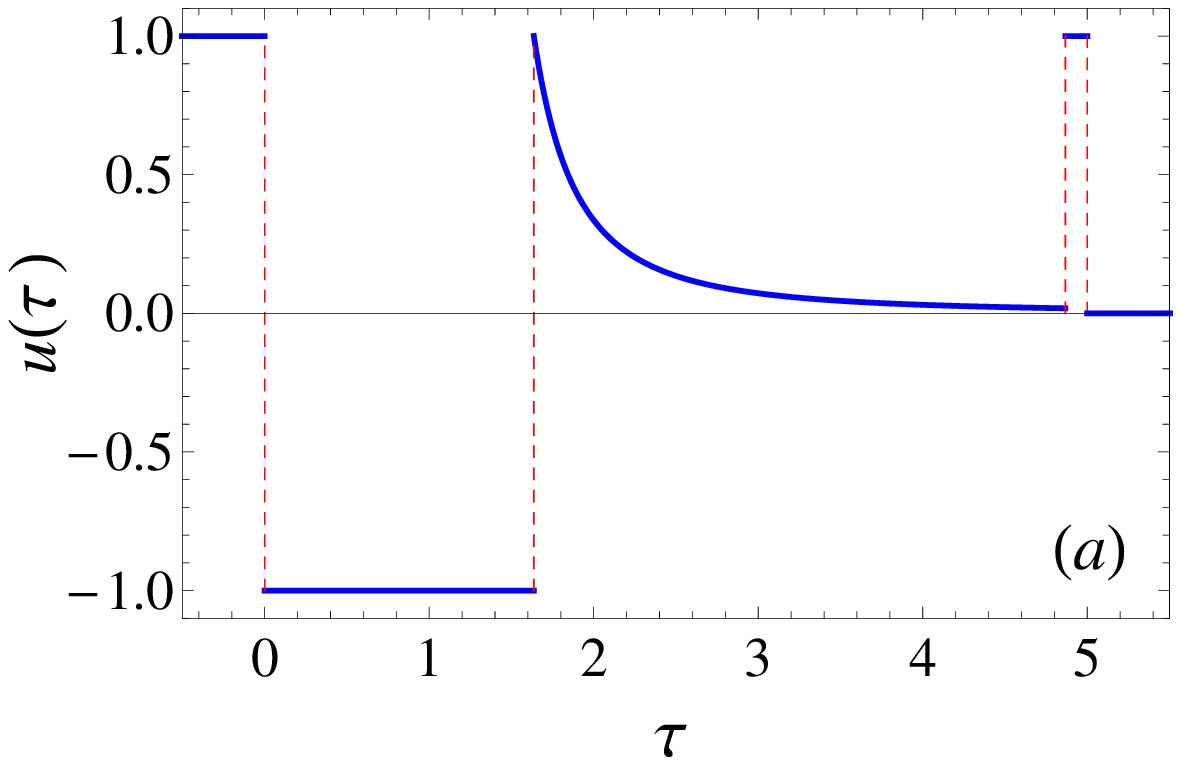}}
\scalebox{0.5}[0.5]{\includegraphics{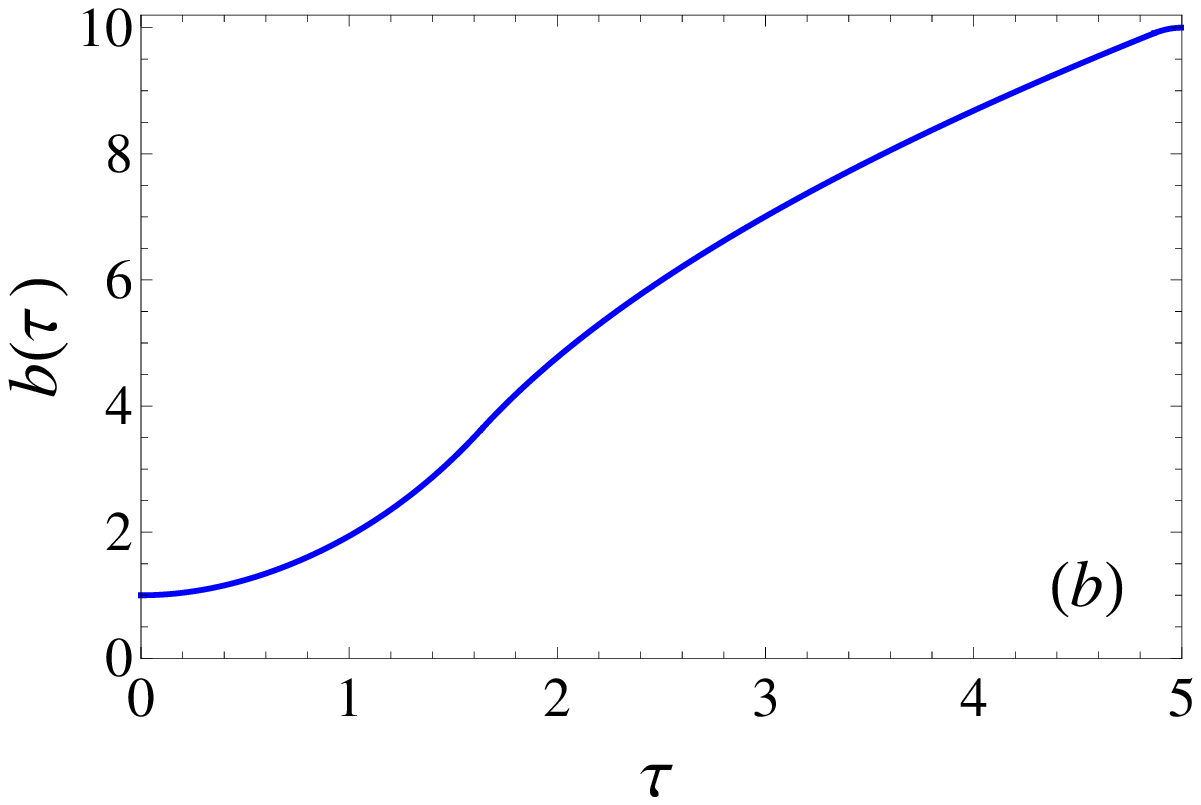}} \caption{(Color
online) (a) Control function $u(t)=\omega^2(t)/\omega^2_0$ versus
$\tau$. (b) Scaling factor $b$ as a function of $\tau$. Parameters:
$\tau_f=5$, $\gamma=10$, and $\delta=1$. } \label{fig3}
\end{figure}

Fig. \ref{fig2} depicts the optimal trajectory of $x_1$ and $x_2$
for  different final times and specific values of $\gamma$ and $\delta$.
The minimal time $\tau_{min}$ is given by the
time-optimal bang-bang solution, $\tau_f$ (bang-bang).
With increasing $\tau_f$, the first and third step times, $\tau_1$
and $\tau_3$,  decrease, and the intermediate step becomes dominant.
Fig. \ref{fig3} shows the control function $u(t)$ and corresponding scaling factor $b$
for bang-singular-bang control when $\tau_f=5$.

At this point the complication of a bang-singular-bang protocol may appear unnecessary, as the bang-bang
control is simpler and takes less time.  However, we shall see in the next section that the bang-bang protocol is also less
stable with respect to the anharmonic perturbation than the bang-singular-bang ones.

%
\subsubsection{Unconstrained frequency}
In the example in the previous section, we set the parameter $\delta=1$, which means that the frequency is bounded
according to  $|u(t)| \leq 1$. If
$\delta$ tends to infinity, the frequency is unconstrained. In this case, Eq.
(\ref{tau}) becomes
\begin{equation}
\tau_1=0,~~\tau_2(c_1)=(\gamma^2-1)/2c_1,~~\tau_3=0,
\end{equation}
so that the final time is $\tau_f=(\gamma^2-1)/2c_1$, $c_2=1$,
and the
trajectory is
\begin{equation}
x_1(\tau)=\sqrt{\frac{\gamma^2-1}{\tau_f}\tau+1},~~\tau\in[0,\tau_f],
\end{equation}
with a scaling factor of the form
\begin{equation}
\label{b-bound}
b(t)=\sqrt{\frac{\gamma^2-1}{t_f}t+1},~~t\in[0,t_f],
\end{equation}
which was found independently using the Euler-Lagrange equation
\cite{Erik} (the boundary conditions (\ref{boundary}) are not completely fulfilled).
The bound for the fidelity, see Eq. (\ref{fb1}), becomes
\beq {{F}}_{EL}= 1 -\left\{\frac{3\hbar}{8m w^2} \left[
t_f+\frac{3(\gamma^2-1)^2}{4 t_f \omega^2_{0}}\right]\right\}.
\label{Fbound}
\eeq
Here and in the next section we rescale time again as $t=\tau/\omega(0)$.
\section{Exact fidelities and anharmonic perturbation energy}
%
%
%
%
%
%
\begin{figure}[ht]
\scalebox{0.5}[0.5]{\includegraphics{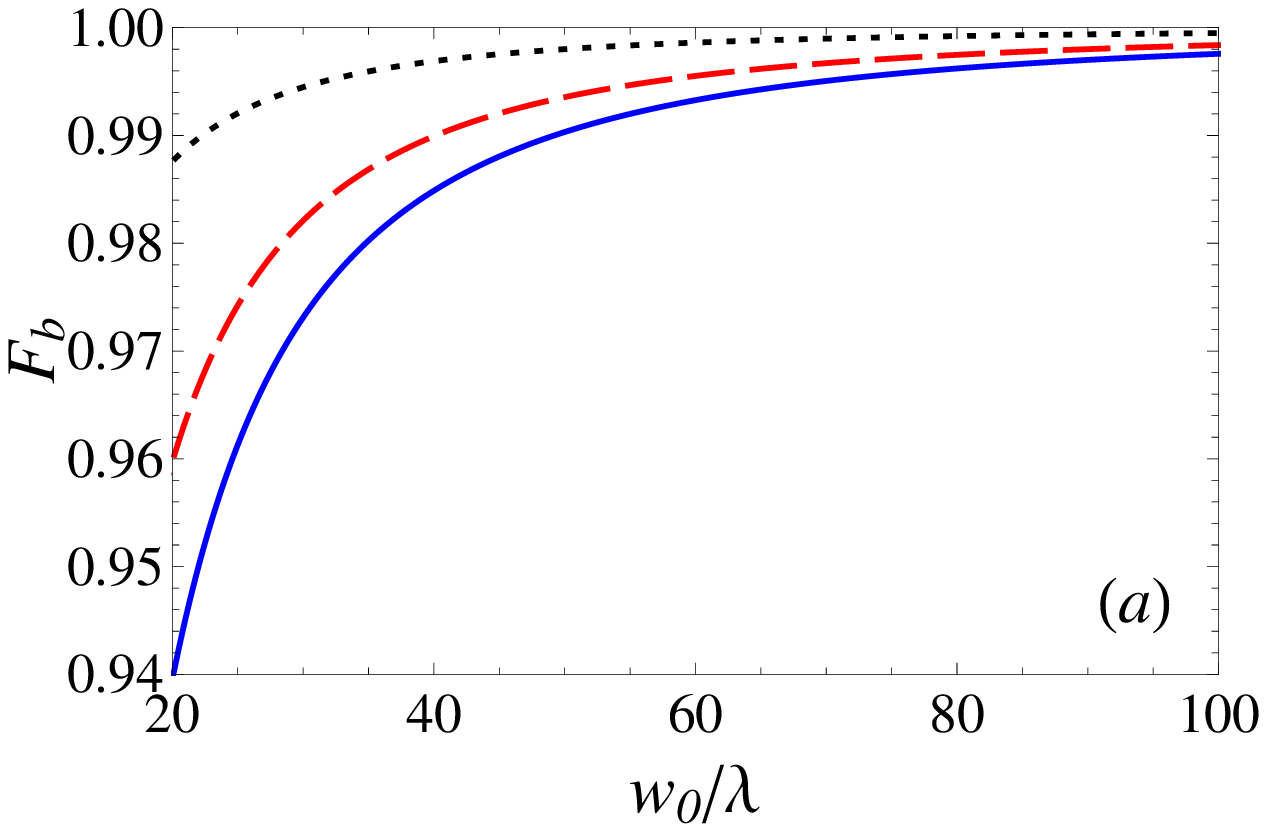}}
\scalebox{0.5}[0.5]{\includegraphics{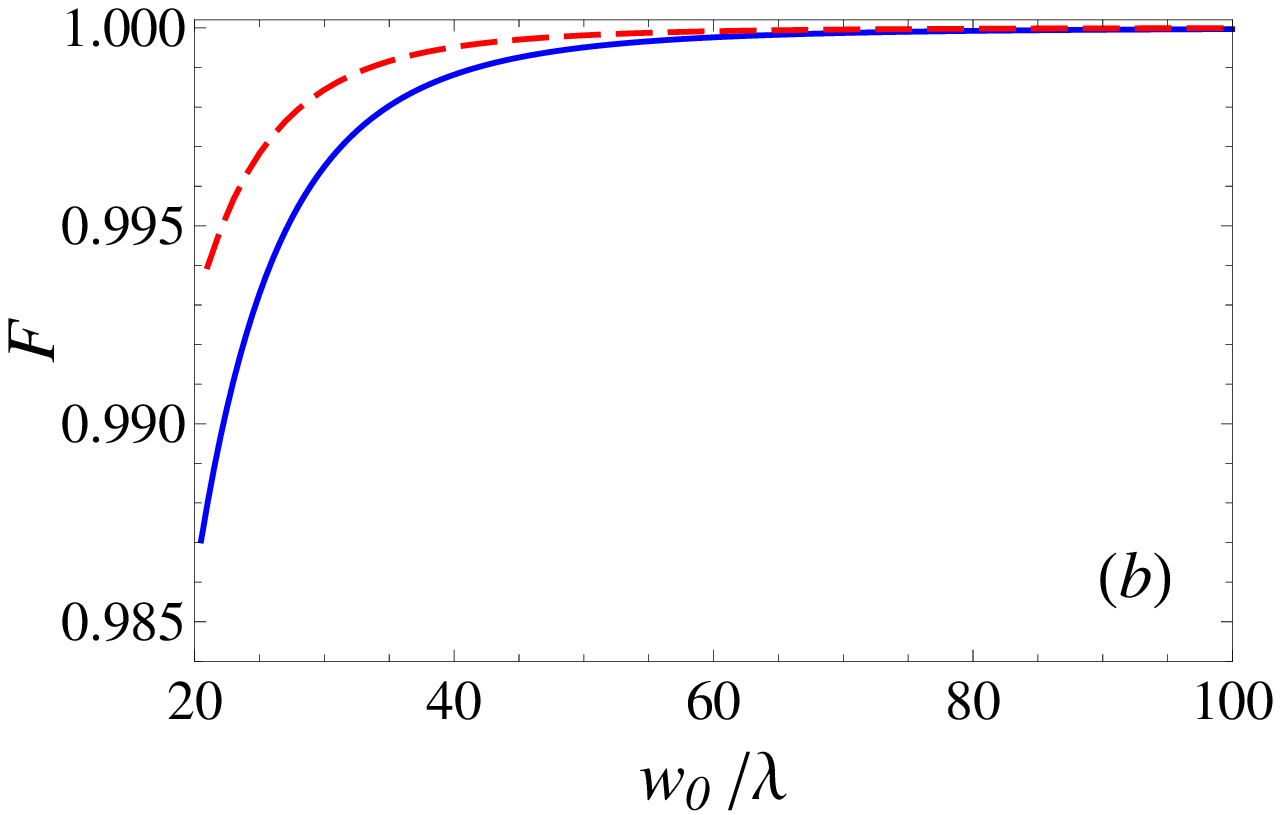}} \caption{(Color
online) Fidelity versus the waist $w$ for the anharmonic trap
(\ref{hami}). (a)  approximate bound $F_b$ in Eq. (\ref{fb2}); (b)
actual fidelity $F=|\langle\psi_0(t_f)| \Psi(t_f)\rangle|$, for
optimal protocols with constrained frequency (dashed red) and
unconstrained frequency (dotted black), and the protocol designed
with  a polynomial ansatz (solid blue). Parameters: $\lambda=1060$
nm, $t_f=0.5$ ms, $\omega(0)=2\pi\times2500$ Hz,
$\omega({t_f})=2\pi\times25$ Hz, $\delta=1$.} \label{fig4}
\end{figure}
\begin{figure}[h]
\scalebox{0.5}[0.5]{\includegraphics{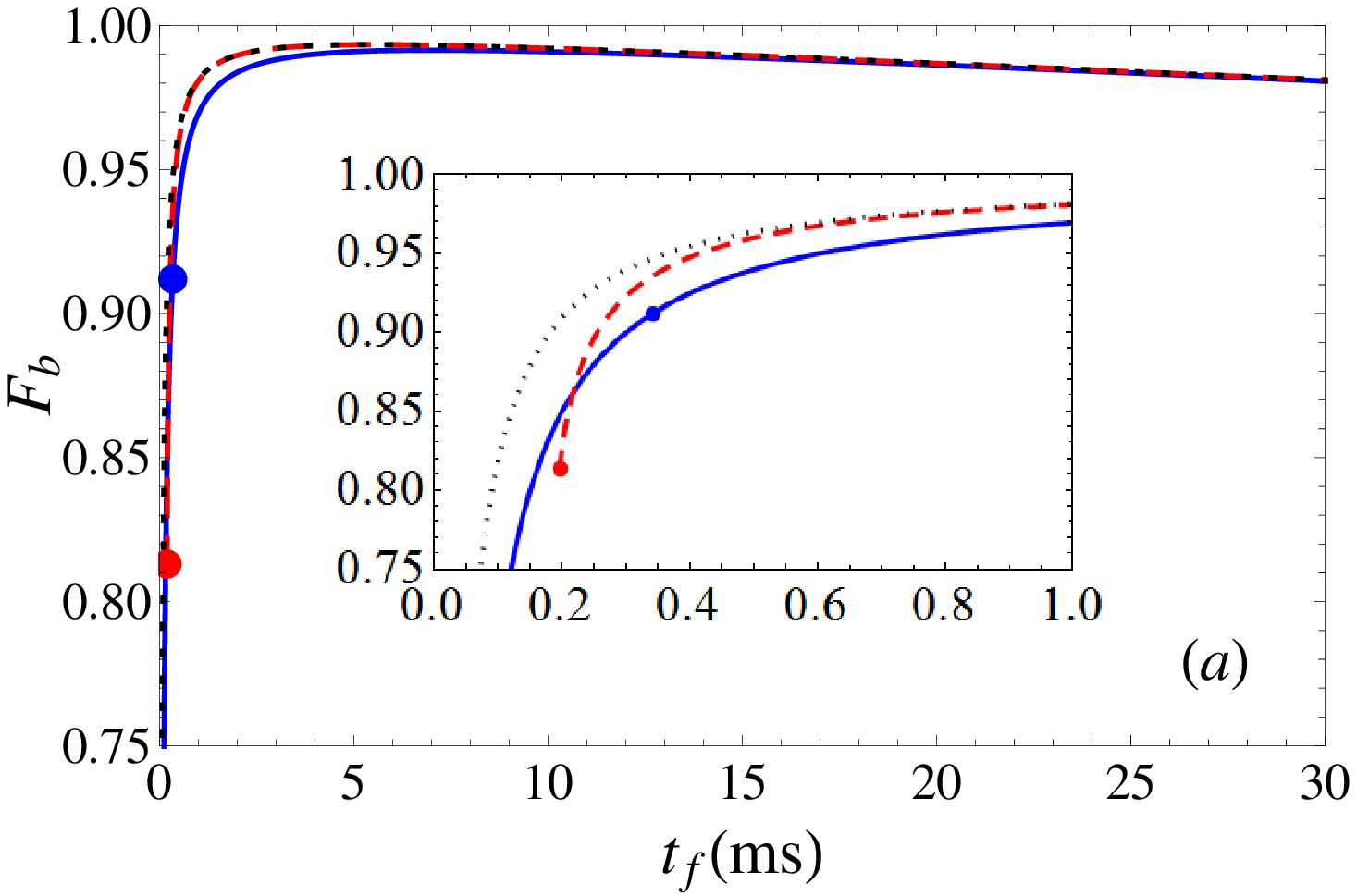}}
\scalebox{0.5}[0.5]{\includegraphics{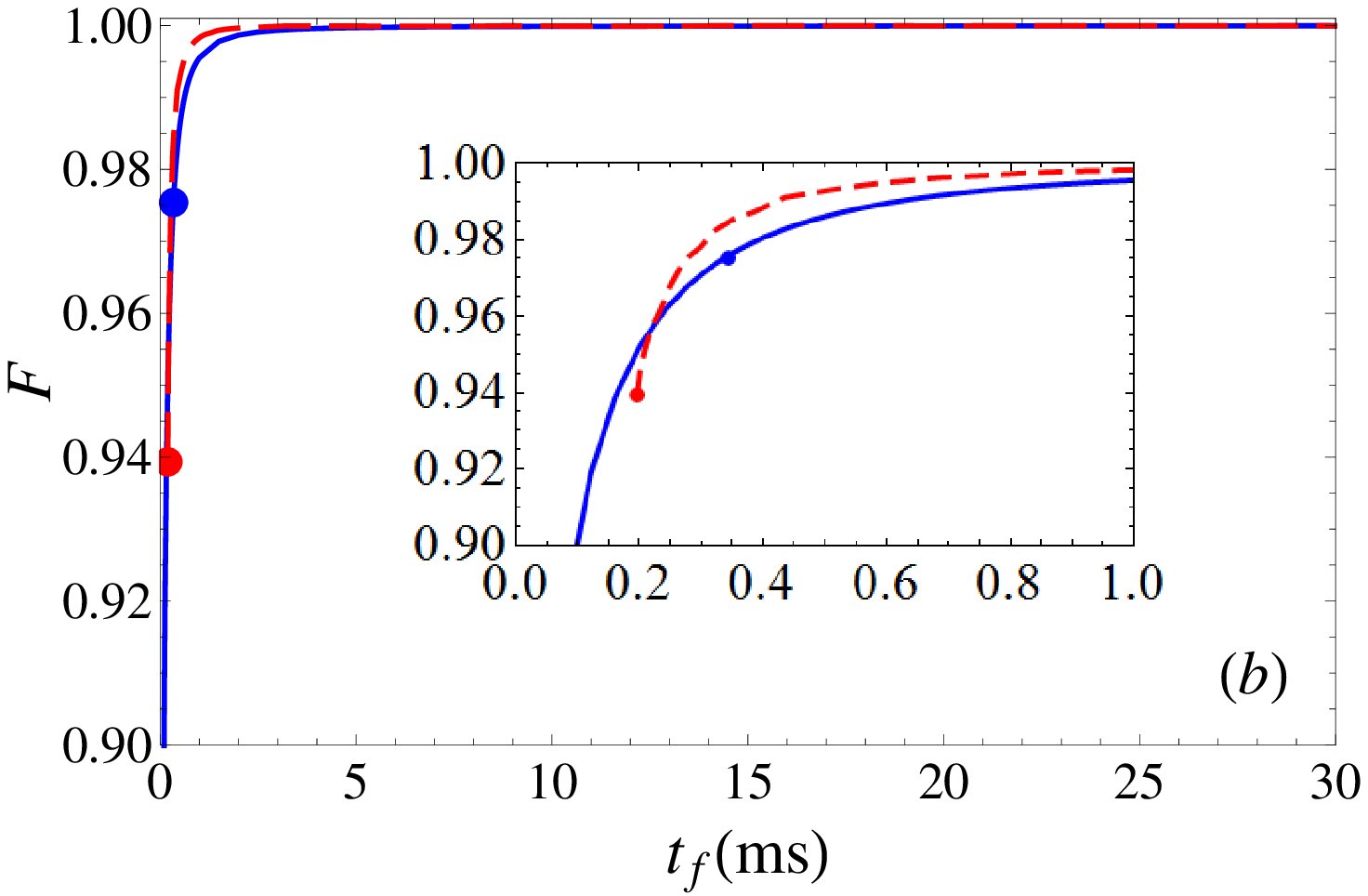}} \caption{(Color
online) Fidelity versus $t_f$  for the  protocol designed with a
polynomial ansatz (solid blue), and for optimal protocols with
constrained frequency (dashed red) and unconstrained frequency
(dotted black). (a) bound $F_b$ in Eq. (\ref{fb2}); (b) actual
fidelity $F=|\langle\psi_0(t_f)| \Psi(t_f)\rangle|$. The insets
amplify the small-$t_f$ region.  The circle on the left end of the
dashed line marks the bang-bang protocol at the minimal time. For
$t_f$ values to the left of the circle on the solid line, the trap
frequency of the polynomial protocol is larger than the bound for
the optimized protocols, at least at some instants. Parameters:
$\lambda=1060$ nm, $w_0=20\lambda$, $\omega(0)=2\pi\times2500$ Hz,
$\omega({t_f})=2\pi\times25$ Hz, and $\delta=1$. } \label{fig5}
\end{figure}

%
So far we have maximized an approximate bound (\ref{fb2}) for the fidelity of a
trap expansion with anharmonic terms. Now we shall calculate for comparison the
actual fidelity, $F=|\langle\psi_0(t_f)| \Psi(t_f)\rangle|$, of the resulting protocol, solving the time-dependent
Schr\"odinger equation with the split-operator method. For
further comparison we also consider the protocol for the pure harmonic
oscillator based on the simple polynomial ansatz $b(t)=\sum_{j=0}^{5}a_j
t^j$ \cite{prl104}. Solving for the coefficients with the boundary
conditions (\ref{boundary}), we get
%
$ b(t)=6(\gamma-1)s^5-15(\gamma-1)s^4+10(\gamma-1)s^3+1, $
%
where $s=t/t_f$. The corresponding frequency is found from the
Ermakov equation (\ref{ermakov}).

Fig. \ref{fig4} (a) shows the (approximate) bound $F_b$ for the polynomial ansatz,
and  for the optimized bang-singular-bang
protocol versus the waist $w_0$ for a fixed $t_f$.
The third (upper) curve is the bound $F_{EL}$ in Eq. (\ref{Fbound}) which,
as the frequency is not constrained for it,  is above the others. 
The actual fidelities,
corresponding to the numerical solution of the time-dependent
Schr\"odinger equation with the designed protocols, are above these
bounds, see Fig. \ref{fig4} (b), which shows the high fidelity achieved
by the bang-singular-bang control.

\begin{figure}[h]
\scalebox{0.5}[0.5]{\includegraphics{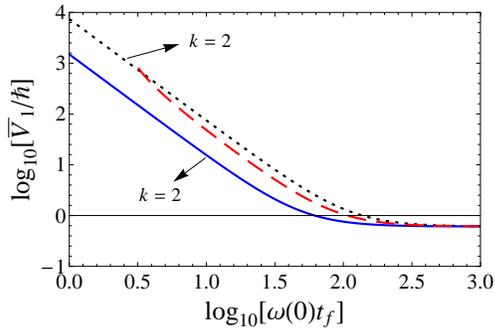}} \caption{(Color
online) Time average of the anharmonic potential energy
$\overline{V}_1/\hbar$ as a function of the final expansion time
$t_f$. $k$ is the scaling exponent. The optimal protocols with
constrained frequency (dashed red) and unconstrained frequency
(dotted black), and the protocol designed with the polynomial ansatz
(solid blue) are compared. Parameters: $w_0=20\lambda$,
$\lambda=1060$ nm, $\omega(0)=2\pi\times2500$ Hz,
$\omega({t_f})=2\pi\times25$ Hz, and $\delta=1$. } \label{fig6}
\end{figure}

Fig. \ref{fig5} depicts the bounds and actual fidelities  with respect to $t_f$ for a
fixed waist $w_0$.
Along the  curve for the optimized protocols (red dashed line), the bang-bang control (the point at the left extreme)
takes the minimal time but is also more sensitive to
the anharmonicity, as it gives the worst fidelity, see Fig. \ref{fig5} (b).
To achieve higher fidelities larger times and thus bang-singular-bang control are necessary.
The fidelity of the optimized  protocol is
higher than that for the polynomial ansatz as long as $|\omega(t)|$  for the polynomial ansatz
stays below the imposed frequency bound for all $t$.  The time $t_f$ below which  this condition does not hold is also marked by a
dot on the polynomial curve. The fidelity bound
$F_b$ decreases after a maximum,
see Fig. \ref{fig5} (a). This behavior is not
reproduced by the actual fidelity, which tends to one as $t_f$
increases. It may be understood by noting that
$F_b=1-\overline{V}_1t_f/\hbar$, where
$\overline{V}_1$
%
is the time-averaged perturbation energy.
From Fig.
\ref{fig6}, $\overline{V}_1\propto t_f^{-2}$ for small $t_f$ but it
tends to a constant value for larger times as no transient
excitations are produced. Correspondingly $F_b$ shows two asymptotic
regimes
\beqa \label{fv} F_b=\left\{
      \begin{array}{ll}
        1-a_1/t_f, & t_f\ll 2\pi/\omega(0) \\
        1-a_2 t_f, & t_f\gg 2\pi/\omega(0),
      \end{array}
    \right.
\eeqa
where $a_1$ and $a_2$ are  constants.

%
%
%
\section{Conclusion}
%
%
%
%
In this work, we have combined invariant-based inverse engineering,
perturbation theory, and OCT to design fast and transitionless
expansions of cold atoms in an anharmonic Gaussian trap. We find that the
optimal protocol obtained from an  approximate fidelity bound $F_b$ is a
bang-singular-bang solution. This protocol minimizes the contribution of the
anharmonicity to the potential energy.

Even though we have specifically treated a one dimensional trap
with the quartic anharmonicity resulting from a Gaussian beam
the results could be applied or generalized
to several other systems presenting anharmonic deviations from a harmonic confinement  in
optomechanics \cite{duan},
mechanical resonators \cite{YongLi}, or trapped ions \cite{SK,Home}.
Ion traps in particular may offer soon the technological possibility to  change the trapping potential on time scales much shorter than the ion oscillation frequencies
facilitating the practical application of bang-bang or bang-singular-bang protocols \cite{Alonso}.
%
%
%
%
\section*{Acknowledgments}
We are grateful to J. Home for useful discussions.
This work was supported by
the National Natural Science Foundation of China
(Grant No. 61176118),
the Shanghai Rising-Star and Pujiang Program (Grant Nos. 12QH1400800 and 13PJ1403000),
the Specialized Research Fund for the Doctoral Program of Higher Education (Grant No. 2013310811003),
the Program for Professor of Special Appointment (Eastern Scholar) at Shanghai Institutions of Higher Learning,
the Basque Government (Grant No. IT472-10), Ministerio de Econom\'\i a y Competitividad (Grant No. FIS2009-12773-C02-01), the UPV/EHU under
program UFI 11/55, and COST programme, under grant number COST-C12.0118.
%


\end{document}